\documentclass[8.5pt,twoside,twocolumn]{article}
\usepackage[super,sort&compress,comma]{natbib}
\usepackage[version=3]{mhchem}
\usepackage{times,mathptmx}
\usepackage{sectsty}
\usepackage{balance}

\usepackage{graphicx} 
\usepackage{amssymb}
\usepackage{lastpage}
\usepackage[format=plain,justification=raggedright,singlelinecheck=false,font=small,labelfont=bf,labelsep=space]{caption}
\usepackage{fancyhdr}

\usepackage{color}

\begin{document}

\thispagestyle{plain}
\fancypagestyle{plain}{
\fancyhead[L]{\includegraphics[height=8pt]{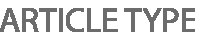}}
\fancyhead[C]{\hspace{-1cm}\includegraphics[height=20pt]{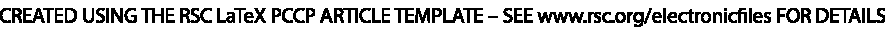}}
\fancyhead[R]{\includegraphics[height=10pt]{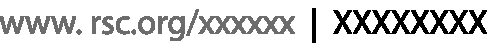}\vspace{-0.2cm}}
\renewcommand{\headrulewidth}{1pt}}
\renewcommand{\thefootnote}{\fnsymbol{footnote}}
\renewcommand\footnoterule{\vspace*{1pt}%
\hrule width 3.4in height 0.4pt \vspace*{5pt}}
\setcounter{secnumdepth}{5}

\newcommand{\medio}[1]{\left\langle #1 \right\rangle}
\newcommand{\erf}{\mathrm{erf}\,}
\newcommand{\erfc}{\mathrm{erfc}\,}
\def\pin{\Pi_{\rm in}}
\def\pout{\Pi_{\rm br}}
\def\peq{\Pi_{\rm eq}}
\def\zetaa{\zeta^*}

\def\pb{p_{\rm bond}}

\def\phigel{\phi_{\rm gel}}
\def\phiglass{\phi_{\rm glass}}

\def\pglass{p_{\rm glass}}
\def\pgel{p_{\rm gel}}

\def\tarrest{\tau_{\rm arrest}}
\def\phiarrest{\phi_{\rm arrest}}
\def\parrest{p_{\rm arrest}}

\def\<{\langle}
\def\>{\rangle}
\def\km{k_{\rm min}}
\def\kM{k_{\rm max}}

\makeatletter
\def\subsubsection{\@startsection{subsubsection}{3}{10pt}{-1.25ex plus -1ex minus -.1ex}{0ex plus 0ex}{\normalsize\bf}}
\def\paragraph{\@startsection{paragraph}{4}{10pt}{-1.25ex plus -1ex minus -.1ex}{0ex plus 0ex}{\normalsize\textit}}
\renewcommand\@biblabel[1]{#1}
\renewcommand\@makefntext[1]%
{\noindent\makebox[0pt][r]{\@thefnmark\,}#1}
\makeatother
\renewcommand{\figurename}{\small{Fig.}~}
\sectionfont{\large}
\subsectionfont{\normalsize}

\fancyfoot{}
\fancyfoot[LO,RE]{\vspace{-7pt}\includegraphics[height=9pt]{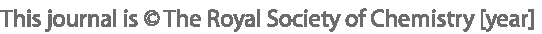}}
\fancyfoot[CO]{\vspace{-7.2pt}\hspace{12.2cm}\includegraphics{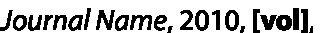}}
\fancyfoot[CE]{\vspace{-7.5pt}\hspace{-13.5cm}\includegraphics{headers/RF}}
\fancyfoot[RO]{\footnotesize{\sffamily{1--\pageref{LastPage} ~\textbar  \hspace{2pt}\thepage}}}
\fancyfoot[LE]{\footnotesize{\sffamily{\thepage~\textbar\hspace{3.45cm} 1--\pageref{LastPage}}}}
\fancyhead{}
\renewcommand{\headrulewidth}{1pt}
\renewcommand{\footrulewidth}{1pt}
\setlength{\arrayrulewidth}{1pt}
\setlength{\columnsep}{6.5mm}
\setlength\bibsep{1pt}

\twocolumn[
  \begin{@twocolumnfalse}
\noindent\LARGE{\textbf{Dynamical arrest: interplay of the glass and of the gel transitions}
\vspace{0.6cm}


\noindent\large{
Nagi Khalil,\textit{$^{a\ddag}$}}, Antonio Coniglio\textit{$^{b}$}, Antonio de Candia\textit{$^{b,d}$}, Annalisa Fierro\textit{$^{b}$}, and Massimo Pica Ciamarra\textit{$^{b,c}$}}\vspace{0.5cm}

\noindent\textit{\small{\textbf{Received Xth XXXXXXXXXX 20XX, Accepted Xth XXXXXXXXX 20XX\newline
First published on the web Xth XXXXXXXXXX 200X}}}

\noindent \textbf{\small{DOI: 10.1039/b000000x}}
\vspace{0.6cm}

\noindent \normalsize{ The structural arrest of a polymeric
suspension might be driven by an increase of the cross--linker
concentration, that drives the gel transition, as well as by an
increase of the polymer density, that induces a glass transition.
These dynamical continuous (gel) and discontinuous (glass)
transitions might interfere, since the glass transition might
occur within the gel phase, and the gel transition might be
induced in a polymer suspension with glassy features. Here we
study the interplay of these transitions by investigating via
event--driven molecular dynamics simulation the relaxation
dynamics of a polymeric suspension as a function of the
cross--linker concentration and the monomer volume fraction. We
show that the slow dynamics within the gel phase is characterized
by a long sub-diffusive regime, which is due both to the crowding
as well as to the presence of a percolating cluster. In this
regime, the transition of structural arrest is found to occur
either along the gel or along the glass line, depending on the
length scale at which the dynamics is probed. Where the two line
meet there is no apparent sign of higher order dynamical
singularity. Logarithmic behavior typical of $A_{3}$ singularity
appear inside the gel phase along the glass transition line. These
findings seem to be related to the results of the mode coupling
theory for the $F_{13}$ schematic model.

} \vspace{0.5cm}
 \end{@twocolumnfalse}
  ]

\footnotetext{\textit{$^{a}$~Departamento de F\'{\i}sica, Universidad de Extremadura, E-06071 Badajoz, Spain}}
\footnotetext{\textit{$^{b}$~CNR--SPIN, Dipartimento di Scienze Fisiche, University of Napoli Federico II, Italy}}
\footnotetext{\textit{$^{c}$~Division of Physics and Applied Physics, School of
Physical and Mathematical Sciences, Nanyang Technological University,
Singapore}}
\footnotetext{\textit{$^{d}$~INFN, Sezione di Napoli,
Complesso Universitario di Monte S. Angelo, Via Cintia, Edificio 6,
80126 Naples, Italy
}}

\section{Introduction\label{sec:introduction}}
A polymeric suspension might behave as a hard sphere system, when
no cross--linkers are added to it, as its dynamics slows down as
the concentration increases. In this case, a transition of
structural arrest, the glass transition, occurs when the volume
fraction $\phi$ overcomes a critical threshold, $\phiglass$. This
transition marks the arrest of the dynamics at all relevant length
scales, for wavevectors ranging from $\km = 2\pi/L$ to $\kM =
2\pi/\sigma$, with $L$ and $\sigma$ sizes of the system and of the
particles. The glass transition is a discontinuous dynamical phase
transition because its order parameter, known as non--ergodicity
parameter and defined as the infinite time limit of relaxation
functions, is zero below the transition but acquires a finite
value at the transition. 
Polymeric suspensions may also undergo a different kind of
structural arrest transition, the gel
transition. This occurs at fixed volume fraction on increasing the
cross--linker concentration, which is related to the probability
that two close monomers are permanently bonded. This concentration
can be tuned, for instance, by radiation as in light induced
polymerization processes (e.g., dental filling pastes) or by heat
(e.g., cooking). On increasing the bonding probability, one drives
the system across a gel transition, where a spanning cluster of
connected monomers emerges. This leads to a gel transition line
$\phigel(p)$ in the $p$--$\phi$ plane. This structural transition
also corresponds to a dynamical arrest transition, as the dynamics
becomes frozen on the smallest wavevector $\km$ being the spanning
cluster unable to diffuse. Contrary to the glass transition, this
is a continuous transition, as the order parameter continuously
increases as one enters the gel phase on increasing $p$ or $\phi$.

The interference of different transitions of dynamical arrest
have been previously investigated, via model coupling theory,
in systems of particles interacting with a hard
core repulsion competing with a very short range attraction.
These systems show an attractive and a repulsive glass
transition, both of them discontinuous, and mode coupling theory
predicts that their interference leads to an
$A_3$ high order singularity
characterized by logarithmic decay of the correlation functions
\cite{dawson,f13,Gotze}.
These predictions were
verified in different colloidal systems~\cite{eckert, pham, chen,
puertas, sciortino, lu}, both experimentally and numerically.
Similar findings were also observed in Ref.~\cite{berthier}, where
the dynamics of a physical gel was numerically studied at high
densities where gelation and glass transition interfere.



\begin{figure}[t!]
\centering
\includegraphics[scale=0.33]{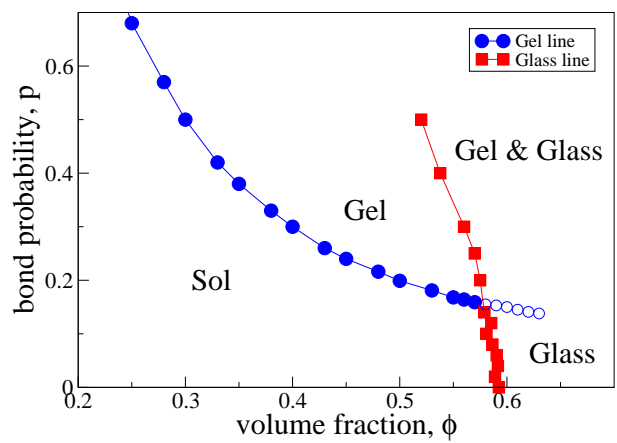}
\caption{\label{fig:percolation}
Structural arrest diagram as a function of the volume fraction,
$\phi$, and of the bonding probability, $p$, illustrating the
interplay of the gel and the glass transition lines. The gel line
is determined via percolative analysis, after introducing bonds
with probability $p$ in an equilibrium (full symbols) or in an
out--of--equilibrium (open symbols) monomer suspension. The glass
line is defined as that where the extrapolated diffusion
coefficient vanishes. Only points up to $p=0.5$ $\phi=0.54$ are
shown, where the self Intermediate Scattering Function (sISF) exhibits a 
logarithmic behavior, which we associate to 
an $A_3$ singularity as predicted  by the $F_{13}$ mode coupling schematic
model (see text and Fig.~\ref{fig:log2}). Solid lines are guides to the eye. }
\end{figure}

In this paper we study the interference of a continuous and
of a discontinuous transition of structural arrest focusing
on a system with permanent bonds where, as
illustrated in Fig.~\ref{fig:percolation}, the gel $\phigel(p)$
and the glass transition line $\phiglass(p)$ intersect.
We address
some questions raised by the presence of these two transition
lines. First, we note that the gel transition corresponds to
structural arrest occurring at the largest ($\km$) scale, while all
scales, including the smallest one ($\kM$), are relevant in the
structural arrest corresponding to the glass transition.
Therefore, we consider the length scale dependence of the
transition of structural arrest. Second, we note that the glass
transition line enters in the gel phase at a point where both
continuous and discontinuous transition occur.
Hence we expect that the model system here studied might have some
analogies with the mode coupling theory for schematic $F_{13}$
model\cite{f13}, which gives two arrested lines: a continuous
transition, which meets a discontinuous one.  This discontinuous
transition ends on a high order critical point ($A_3$ singularity)
characterized by logarithm decay\cite{Gotze,f13,dawson} of the
relaxation functions. Topologically the phase diagram is similar
to the one found here (Fig.~\ref{fig:percolation}). Consistently
with mode coupling theory, we find logarithmic behavior at a point well in the 
gel phase along the glass line.


\section{Numerical model}
\label{sec2}
We perform event driven molecular dynamics simulations~\cite{lu91,alti87}
of a $50{:}50$ binary mixture of $N = 10^3$ hard spheres (monomers) of
mass $m$ and diameters $\sigma$ and $1.4\sigma$,
in a box of size $L$ with periodic boundary conditions.
The chosen size ratio is known to effectively prevent crystallization
\cite{chbesa10,bewi09,xuhalina09,ohlalina02}.
The volume fraction $\phi = Nv/L^3$, where $v$ is the average particle volume,
is tuned by changing the size $L$ of the box. The mass $m$, the diameter $\sigma$ of the smaller particles,
and the temperature $T$, fix our mass, length and energy scales, while the time unit is $\sqrt{m\sigma^2/T}$.
After thermal equilibration at the desired volume
fraction, permanent bonds are introduced with probability $p$
between any pair of particles separated by less than $1.5\sigma$.
A bond corresponds to an infinite square well potential, extending from
$\sigma$ to $1.5\sigma$.
The procedure we use to insert the bonds mimics
a light--induced polymerization process,
as the number of bonds depends on both $p$ and $\phi$.

Using the percolation approach, we identify the gel phase as the state in which a percolating
cluster is present \cite{flory,degennes}, and the gelation transition as the percolation line.
A standard finite--size scaling analysis of the mean
cluster size~\cite{stauffer} is applied to identify the percolation line.
The resulting line is illustrated
in Fig.~\ref{fig:percolation}.
The gel transition occurs at a critical bond probability
$\pgel(\phi)$ decreasing with $\phi$.

\section{Standard glass and gel behaviours}
\begin{figure}[t!]
\centering
\includegraphics[scale=0.35]{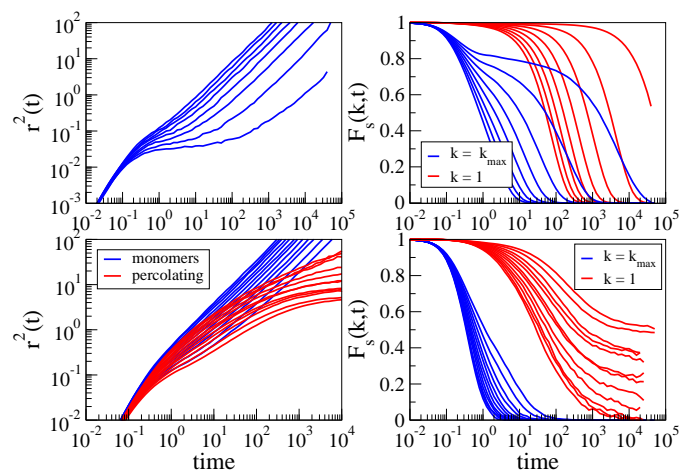}
\caption{\label{fig:glassgel}
Mean square displacement (left) and sISF (right).
Top panel: on approaching the glass
transition at $p = 0$, with volume
fraction from $\phi = 0.52$ to $\phi = 0.59$ (from left to right).
Bottom panel: on entering the gel phase at $p = 0.3$ ($\phigel \approx 0.4$),
with volume fraction from $\phi = 0.4$ to $\phi = 0.51$ (from left to right).
}
\end{figure}

Since we are interested in the interplay of the gel and of the glass transition lines,
we start by shortly reviewing the main features of the slowing down of the dynamics
when these transitions do not interfere.
First, we consider the transition at zero bonding probability, where
our system reduces to a hard sphere suspension and undergoes a
glass transition as the volume fraction increases.
This crowding induced dynamical transition is characterized
by well known signatures in the mean square displacement, $r^2(t)$, and in
the self intermediate scattering function (sISF), $F_s(k,t)$, defined respectively by:
\begin{equation}
 \<r^2(t)\> = \frac{2}{N}\sum_{i=1}^{N/2} r_i^2(t),
\label{msd}
\end{equation}
and
\begin{equation}
 F_s(k,t) = \frac{2}{N} \sum_{i=1}^{N/2} e^{i k (r_i(t)-r_i(0))},
\label{self}
\end{equation}
where the sums extend only to the larger particles.
These are reviewed in the top row of Fig.~\ref{fig:glassgel}.
At high densities, particles rattle in the cages formed by
their neighbours before entering the diffusive regime.
Accordingly, the mean square displacement
develops a plateau at intermediate times, known as Debye--Weller factor.
This plateau becomes longer and longer as the density increases.
The diffusivity $D = \lim_{t \to \infty} r^2(t)/6t$ decreases on
increasing $\phi$, and vanishes at the glass transition.
Consistently with the behaviour of $r^2(t)$,
the relaxation function at large $k$ develops
a two step decay, the first one ($\beta$ relaxation)
associated to the rattling within the cages,
the second one ($\alpha$) to the onset
of diffusive motion.
The $\beta$ relaxation becomes less and less visible
as $k$ decreases, and length scales
larger than those associated to the typical vibrational motion are probed.
At the glass transition, only the $\beta$ relaxation occurs,
and $F_s(k,t)$ asymptotically reaches a finite value,
known as non--ergodicity parameter $f_k$.
The glass transition is a discontinuous phase transition
as $f_k$ varies discontinuously across the transition.

We now consider the transition at high $p$, where on increasing $\phi$ the system undergoes a gel transition
at a volume fraction $\phigel \ll \phiglass$, so that the two transitions do not interfere.
The typical behaviour of $r^2(t)$ and of $F_s(k,t)$ on approaching this transition
is illustrated in Fig.~\ref{fig:glassgel} (bottom row).
Due to the presence of bonds, in the system there are clusters with a different size $s$.
In the long time limit, each cluster diffuses
with a diffusivity decreasing with $s$. Accordingly, at long times
the mean square displacement is essentially fixed by that of the fastest clusters, the monomers.
Since the volume fraction is small, the diffusivity of the monomers is only slightly affected
by the volume fraction and has not special features at the gel transition,
above which the monomers diffuse within the percolating polymer network.
On the contrary the largest cluster, whose mean square displacement is
also illustrated in Fig.~\ref{fig:glassgel}, slows down
on approaching the gel transition, and indeed its diffusivity vanishes
at the transition (where the largest cluster percolates).
We note that, on increasing $\phi$, the mean square displacement of
the largest cluster does not reach the diffusive regime on the time scale
of our simulations.  Indeed, it has three contributions: the first is due
to the vibrations of the particles in the cages formed by nearest neighbors, the second to the fluctuations of the cluster structure,
and the third to the diffusion of its center of mass. Of these, only the third reaches the diffusive regime at long times.
At the gel transition, when the diffusion of the center of mass of the largest
cluster vanishes, the mean square displacement remains subdiffusive at all times, while inside the gel it eventually reaches a plateau.

The behaviour of the sISF
critically depends on the length scale at which it is probed. At
small scales, that is for $k = \kM$, it does not exhibit any
particular feature on approaching the gel transition. On the other
hand, at larger length scales, $k = \km$, the relaxation time
diverges when $\phi$ approaches $\phigel(p)$. At the transition,
the correlator decays in time as a power law, while inside the gel
asymptotically tends to a plateau, $f_{k}$, the non--ergodicity
parameter, that is zero at the the transition, and
continuously grows on entering the gel phase. The growth is
continuous as the value of $f_{k}$ is related to the density of
the particles in the percolating cluster, and of the monomers that are
trapped in the percolating cluster. The gel transition therefore
induces a continuous dynamical phase transition. These properties
are found in gelling systems both experimentally
\cite{gel1,gel2,gel3} and numerically \cite{ema,fene,fene-lungo,jcp}.
In Ref.s \cite{zippelius,fierro}, deep analogies of gelation
and spin glass transition was suggested.

\section{Dynamics in the gel phase}
We now consider how the features of the relaxation process changes as the volume fraction increases,
and the system enters the gel phase and approaches the glass transition. This is
identified as the volume fraction where the extrapolated diffusivity of the monomers vanishes (see Fig.~\ref{fig:percolation}).
In this section we focus on a high enough value of the bond probability, $p = 0.3$, for which
$\phigel \approx 0.4$
is sensible smaller than $\phiglass \approx 0.58$.

\subsection{Mean square displacement}
\begin{figure}[t!!!]
\centering
\includegraphics[scale=0.33]{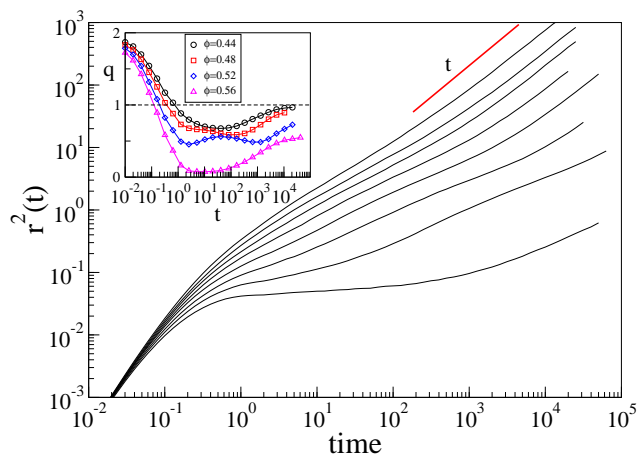}
\caption{\label{fig:msdinside}
Mean square displacement in the gel phase, for $p = 0.3$ and $\phi$ from
$0.42$ to $0.56$ (from top to bottom).
In the inset, the subdiffusive exponent $q=d\log r^2(t)/d\log t$,
as function of time for $p = 0.3$ and $\phi=0.44,~0.48,~0.52,~0.56$.
}
\end{figure}
\begin{figure}[t!!]
\centering
\includegraphics[scale=0.33]{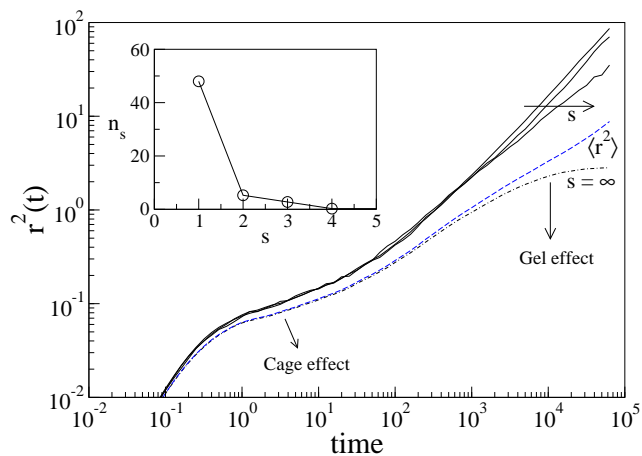}
\caption{\label{fig:msdclusters}
Main panel: mean square displacement averaged over all the particles, $\<r^2\>$,
over particles belonging to finite clusters of size $s = 1$, $2$, $3$,
and over particles belonging to the percolating clusters, for $p = 0.3$ and $\phi = 0.54$.
Inset: size distribution of finite clusters for the same values of $p$ and $\phi$.
There is also a percolating cluster, with $s \approx 930$ particles.}
\end{figure}

In the gel phase, the mean square displacement at intermediate
times shows a subdiffusive regime, similar to that observed in the slow dynamics of
colloidal particles diffusing in a matrix of disordered hard sphere obstacles \cite{kurzidim}.
We have characterized it by investigating the time dependence of the
subdiffusive exponent $q$, defined by $q=d\log r^2(t)/d\log t$.
Data are illustrated in Fig.~\ref{fig:msdinside}.
This exponent exhibits a crossover from the ballistic short time value $q = 2$, to the
diffusive long time value, $q = 1$.
At intermediate times, a subdiffusive
behaviour is found, $q < 1$, which becomes prominent as the volume fraction increases.
Further increasing the volume fraction, very peculiar features appear,
as the exponent $q$ might also exhibit two minima.
This behaviour is rationalized considering that the mean square displacement, Eq. (\ref{msd}),
has contributions from particles belonging to different finite clusters
of size $s$, as well as from the spanning cluster ($s = \infty$, in the thermodynamic limit), as
\begin{equation}
 \<r^2(t)\> = \frac{1}{N} \sum_s s n_s \<r_s^2(t)\>,
\end{equation}
where the sum runs over the cluster size $s$, $\<r_s^2(t)\>$ is the mean square displacement of
clusters of size $s$, and $n_s$ is the number
of cluster with size $s$. The cluster size distribution $n_s$ and the mean
square displacement for different $s$ are illustrated in Fig.~\ref{fig:msdclusters}, for $p = 0.3$ and
$\phi = 0.54$. The figure clarifies that two phenomena lead to the subdiffusive
behaviour. The first one is the crowding induced cage effect typical of glasses,
which involves both particles belonging to finite clusters, as well as particles belonging to
the percolating cluster. The second one only involves the particles of the percolating cluster,
whose mean square displacement reaches a plateau at long times, as the percolating cluster is unable to diffuse.
The coexistence of these two effects, whose relevant importance is fixed by $\phi$ and $p$,
leads to the unusual features of the mean square displacement plotted in Fig.~\ref{fig:msdinside}.

\begin{figure}[t!!]
\centering
\includegraphics[scale=0.33]{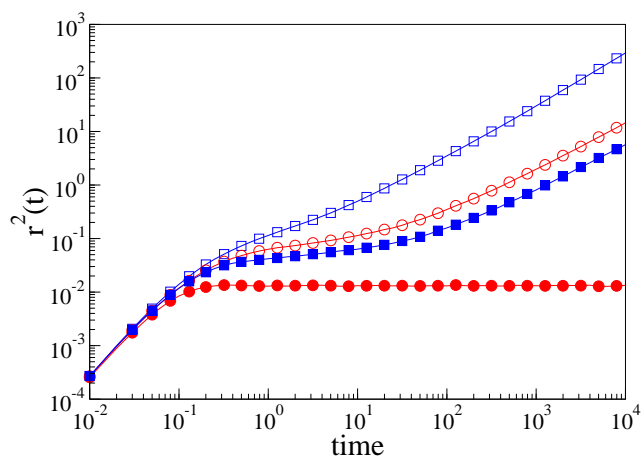}
\caption{\label{fig:msd_bloccato}
The mean square displacement of the monomers in the gel (open symbols)
is compared with that obtained when the spanning cluster is frozen (full symbols).
Squares: $p = 0.2$, $\phi = 0.52$; Circles: $p = 0.25$, $\phi = 0.55$.
}
\end{figure}

As we have observed above, the diffusive problem in the gel phase
shares some similarities
with the much investigated problem of single--particle diffusion
in a matrix of disordered hard sphere obstacles \cite{lorentzgas, kurzidim, nicolai},
where single particles are considered to move in a frozen environment.
Conversely, in the present case, monomers and finite clusters diffuse
in a fluctuating environment, as the percolating clusters may vibrate.
These fluctuations affect the diffusion of free particles as we see in
Fig.~\ref{fig:msd_bloccato},
where the mean square displacement
obtained in the actual gel is compared with that obtained by freezing the percolating cluster.
The figure clarifies that the fluctuations of the percolating cluster strongly influence the diffusivity of the
particles. For instance, at $p = 0.25$, $\phi = 0.55$, the system is completely arrested by
freezing the percolating cluster, showing that the relaxation of the system is completely due to the fluctuations of the percolating cluster.
This result is relevant to model diffusion in
biological systems, e.g. to investigate drug delivery processes.

\subsection{Correlation functions}
In this section, we describe the relaxation of the system in the gel phase
through the study of the sISF, Eq. (\ref{self}).
Conventionally, the relaxation time $\tau_k$ of the system is defined
requiring that $F_s(k,\tau_k) \equiv 1/e$, or a similar small value,
and the transition of structural arrest is considered to occur
when $\tau_k$ reaches a high value $\tarrest$, related the experimental time scale.
Here, we fix $\tarrest\equiv 10^5$,
and explicitly consider the dependence on the threshold
used to define the relaxation time. Therefore, we derive a $k$ and $x$
dependent structural arrest line $\phiarrest(p)$, defined as the volume fraction at which $F_s(k,\tarrest) \equiv x$.

\begin{figure}[t!!]
\centering
\includegraphics[scale=0.33]{phi_arrest_x.eps}
\includegraphics[scale=0.82]{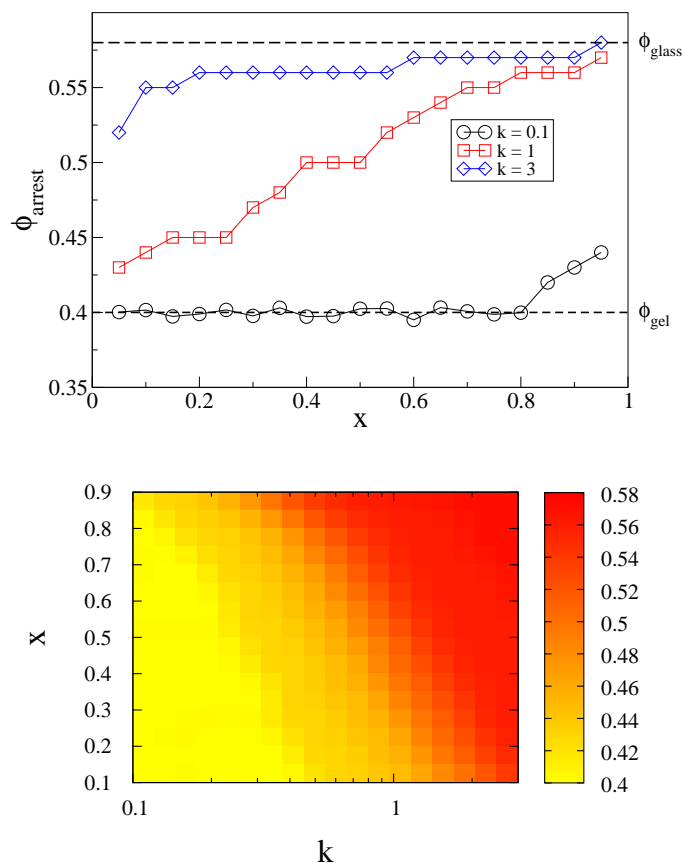}
\caption{\label{fig:phikx}
We consider the transition of structural arrest to occur when
$F_s(k,\tarrest) \equiv x$, with $\tarrest \equiv 10^5$.
The top panel illustrates the dependence of $\phiarrest$ on $x$,
for three values of $k$, and $p = 0.3$.
The bottom panel is a colour map showing the dependence of
$\phiarrest$ on $k$,  for $p = 0.3$.
The figures clarify that at small $k$, $\phiarrest \approx \phigel$,
at large $k$, $\phiarrest \approx \phiglass$,
while at intermediate $k$ $\phiarrest$ exhibits a crossover between these
two limiting values on increasing $x$.}
\end{figure}

Fig.~\ref{fig:phikx} (top panel) illustrates the dependence of
$\phiarrest$ on $x$ for three different values of $k$,
at $p = 0.3$. The figure clarifies the existence of
two characteristic values for
$\phiarrest$, $\phigel \approx 0.4$ and $\phiglass \approx 0.58$.
Indeed, regardless of $x$, $\phiarrest \approx \phigel$
at small $k$, and $\phiarrest \approx \phiglass$ at large $k$.
At intermediate values of $k$, $\phiarrest$ exhibits a crossover
from $\phigel$ to $\phiglass$ as $x$ increases.
The decay of the correlation function
at wavevectors $\km < k < \kM$ is therefore influenced by both
the gel and the glass transitions.
The bottom panel of Fig.~\ref{fig:phikx} illustrates the
same crossover for different wavevectors,
that influence the value of $x$ at which the crossover occurs.
The glassy time scale affects the initial decay of the correlation function,
and it is thus observed for large $x$, while the gel relaxation
corresponds to the final relaxation, and it is observed for small $x$.

\begin{figure}[t!]
\centering
\includegraphics[scale=0.33]{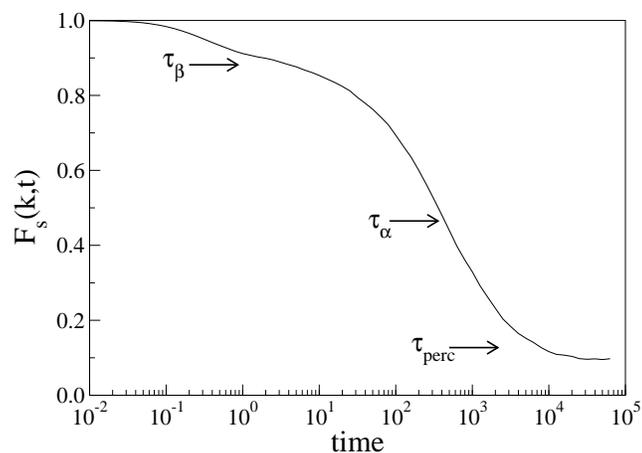}
\caption{\label{fig:new}
The sISF, $F_s(k,t)$, for $k=3$, $p=0.3$, and $\phi=0.54$, larger than
$\phi_{gel} \approx 0.4$.
The three relaxation processes are clearly distinguishable.}
\end{figure}

Data shown in Fig.~\ref{fig:new} further clarify these findings. As also found in Ref. \cite{berthier},
we can recognize three different relaxation time scales:
$\tau_\beta$, due to the rattling of particles in the nearest neighbour cage, does not diverge at all;
$\tau_\alpha$, due to the opening of the cage, diverging at the glass transition line;
and finally $\tau_{perc}$, due to the relaxation of the largest cluster, diverging at the gel transition line
(data plotted in Fig.~\ref{fig:new} refer to $\phi > \phi_{gel}$, then 
$F_s(k,t)$ does not relax to zero, and
reaches at long time a finite value).

\section{Interference of structural arrest lines}

Here, we consider the interference of the glass and of the gel
transitions in a model of a chemical gel, where bonds are
permanent. As we have noticed, the glass transition is a
discontinuous transition and the gel transition is a continuous
one. Hence, we expect that the model system here studied might
have some analogies with the mode coupling theory for schematic
$F_{13}$ model \cite{f13},
which gives two arrested lines: a
continuous transition, which meets a discontinuous one.
The model predict that the discontinuous transition ends on a high order critical point
($A_3$ singularity), where the plateau of the two transitions
coincide, characterized by logarithm decay \cite{Gotze,f13,dawson}
of the relaxation functions.
Topologically the phase diagram is similar to the one found here
(Fig. \ref{fig:percolation}).
\begin{figure}[t!]
\centering
\includegraphics[scale=0.33]{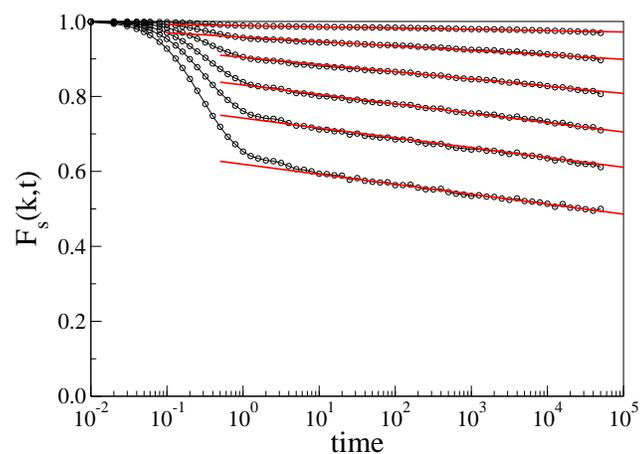}
\caption{\label{fig:log2}
The sISF, $F_s(k,t)$, for $p=0.5$, $\phi=0.52$,
and wavevector $k=1$, $2$, $3$, $4$, $5$, $6.28$ (from top to bottom).
Continuous lines are logarithmic functions.
}
\end{figure}

\begin{figure}[t!]
\centering
\includegraphics[scale=0.33]{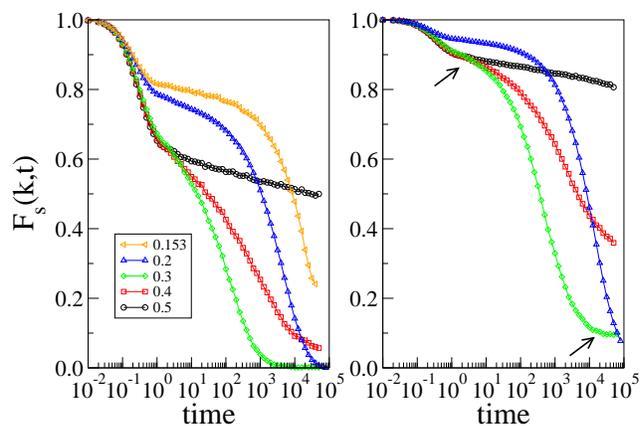}
\caption{\label{fig:selfalong}
The sISF, $F_s(k,t)$, at different values of $p$ as indicated,
moving along the glass line, for $k = \kM$ (left panel) and
for $k = 3$ (right panel).
The plateau associated with the gel and the glass are indicated by an arrow.
}
\end{figure}

In our model, due to long relaxation
time involved, it is rather difficult to localize  such
singularity. However we do find evidence of a logarithmic decay in
a region inside the gel phase, close to the glass transition line,
as shown in Fig.~\ref{fig:log2}.
In order to appreciate how this logarithmic decay emerges, we
follow the evolution of the sISF on increasing $p$ along the glass
line, in the gel phase.
Fig.~\ref{fig:selfalong} shows that on increasing $p$ the
value of the plateau associated to cage motion of the particles
decreases, and the plateau becomes shorter. On increasing
$p$ we also observe the value of the plateau associated to the gel
transition to grow, and the time needed to reach it to increase.
Accordingly, the higher order critical point where a logarithm
decay is found seems to be associated to the state where the two
plateau coincide, in agreement with the $A_3$ singularity of the
$F_{13}$ model.
We stress that the high order
singularity may occur at higher values of $p$ including possibly
$p=1$. Further investigation is necessary.


\section{Conclusions}
The gel and the glass transitions are characterized by different
dynamical signatures, reflecting the fact that the gel transition
is a continuous transition marking the arrest of the dynamics on
the largest length scale of the system, while the glass transition
is a discontinuous transition marking the arrest of the dynamics
at all length scales. This makes of interest the investigation of
the interplay of these transitions in polymeric suspensions, where
the transition are driven by an increase of the cross--linker concentration and the polymer volume
fraction. Here, we have performed
such an investigation, focusing on the glassy dynamics within the
gel phase and on their mutual interference.

An important open question ahead regards the relevance of the mode
coupling theory for schematic $F_{13}$ model~\cite{f13} for these
systems. Our results support this scenario, as we do not only
recover the interference of a continuous and of a discontinuous
transition, but we also find a point where the relaxation function
seems to decay logarithmically, which should be associated to
the A3 singularity. However, more work is needed in this
direction, particularly to better localize this high order
critical singularity.

\bigskip
\noindent{{\bf Acknowledgement}\\
The authors would like to thank Dr. M. Sellitto for having drawn their attention on
the properties of the $F_{13}$ model.
}

\footnotesize{
\bibliographystyle{rsc}

}

\end{document}